\documentclass[twocolumn,showpacs,preprintnumbers,amsmath,amssymb,aps,prl,superscriptaddress]{revtex4-1}
\usepackage{color}
\usepackage{graphicx}
\usepackage{textcomp} 
\usepackage{subeqnarray}
\usepackage[caption=false]{subfig} 
\begin{document}
\title{Topological defects in cholesteric liquid crystal shells}
\author{Alexandre  Darmon}
\affiliation{EC2M, UMR CNRS 7083 Gulliver, ESPCI, PSL Research University, 10 rue Vauquelin, 75005 Paris, France}
\author{Michael Benzaquen}
\affiliation{EC2M, UMR CNRS 7083 Gulliver, ESPCI, PSL Research University, 10 rue Vauquelin, 75005 Paris, France}
\affiliation{Current address: Capital Fund Management, 23 rue de l'Universit\'e, 75007 Paris, France}
\author{Simon \v{C}opar}
\affiliation{Faculty of Mathematics and Physics, University of Ljubljana, Jadranska 19, 1000 Ljubljana, Slovenia.}
\author{Olivier Dauchot}
\affiliation{EC2M, UMR CNRS 7083 Gulliver, ESPCI, PSL Research University, 10 rue Vauquelin, 75005 Paris, France}
\author{Teresa Lopez-Leon}
\affiliation{EC2M, UMR CNRS 7083 Gulliver, ESPCI, PSL Research University, 10 rue Vauquelin, 75005 Paris, France}

\begin{abstract}
{We investigate experimentally and numerically the defect configurations emerging when a cholesteric liquid crystal is confined to a spherical shell. We uncover a rich scenario of defect configurations, some of them non-existent in nematic shells, where new types of defects are stabilized by the helical ordering of the liquid crystal. In contrast to nematic shells, here defects are not simple singular points or lines, but have a large structured core. Specifically, we observe five different types of cholesteric shells. We study the statistical distribution of the different types of shells as a function of the two relevant geometrical dimensionless parameters of the system. By playing with these parameters, we are able to induce transitions between different types of shells. These transitions involve interesting topological transformations in which the defects recombine to form new structures. Surprisingly, the defects do not approach each other by taking the shorter distance route (geodesic), but by following intricate paths.}
\end{abstract}
\maketitle


\section{Introduction}

Liquid crystal droplets have been extensively studied, both from theoretical and experimental points of view \cite{deGennes1993,Dubois-Violette1969,Williams1986,Bezic1992,Lavrentovich1998,Lopez-Leon2011b,Orlova2015}. They are of particular interest to the scientific community because they represent one of the simplest systems in which topological defects are found to be stable. Indeed, the natural curvature of the spherical interface induces geometrical frustration in the molecular arrangement, resulting in disordered regions called topological defects. These defects are not only interesting from a fundamental point of view, but also control the mechanical and optical properties of the droplet. Many industrial applications have benefited from this interesting feature  \cite{Bahadur}. Switchable windows, in which the optical properties of nematic droplets are tuned by an externally applied electric field, are a good example of this  \cite{Drzaic1986,Drzaic1988}.

The richness in defect configurations increases immensely if a chiral nematic or cholesteric is used to make the droplets. Although chiral nematics in confined geometries have been quite extensively studied in the past  \cite{Bouligand1970,Bouligand1984,Bezic1992,Xu1997,Lavrentovich1998}, state-of-the-art experimental and numerical techniques have revealed a plethora of interesting new structures \cite{Sec2012,Sec2014,Orlova2015,PosnjakG_SciRep6_2016} and possible applications \cite{Lin2011,Geng2013,MannaU_AngewChemIntEd52_2013,AguirreLE_ProcNatlAcadSci113_2016,ZhouY_ACSNano_2016}. In particular, the recent discovery of lasing properties in cholesteric droplets has revived the research in the domain \cite{Humar2010,GardinerDJ_OptExpress19_2011,Uchida2013, Napoli2013, ChenL_AdvOptMater2_2014, Uchida2015, WandCR_PhysRevE91_2015}. Due to molecular chirality, cholesteric liquid crystals display a mesoscopic helical organization, with a repeated distance set by the helical pitch. This layered structure makes each droplet a Bragg resonator, where light emission can be stimulated by including additional dye molecules in the liquid crystal. Such a configuration has an associated topological defect that spans the droplet radius and plays a determinant role in the droplet optical properties. Numerical simulations have provided a detailed description of the molecular organization within the droplet, revealing the intricate double-helix structure of the radial defect  \cite{Sec2012}.

The detailed structure of the double-helix radial defect has first been observed experimentally in water-cholesteric-water double emulsions  \cite{Darmon2015}. In this geometry, the liquid crystal is not confined to a bulk droplet, but to a thick spherical shell. This configuration enables tuning the chirality of the system by playing with the shell thickness-to-pitch ratio.  At high chirality, the shell displays a radial defect with an intricate double-helix structure, as predicted by simulations for a bulk cholesteric droplet. However, at low chirality, the shell is characterized by two defects, each of them made of a number singular rings that piles up with a certain separation distance. Between these two limit cases, new defect configurations are expected to emerge  \cite{WandCR_PhysRevE91_2015}. These new configurations might be relevant in the context of optical applications  \cite{Uchida2013, ChenL_AdvOptMater2_2014,Uchida2015,Lagerwall2016} and in the design of new building blocks for colloidal self-assembly  \cite{Poon2004,Lavrentovich2011,Geng2013,Lopez-Leon2011,Sec2012b, Lagerwall2012, Koning2013}.

\begin{figure*}[t!]
 \centering
 \includegraphics[width=1\textwidth]{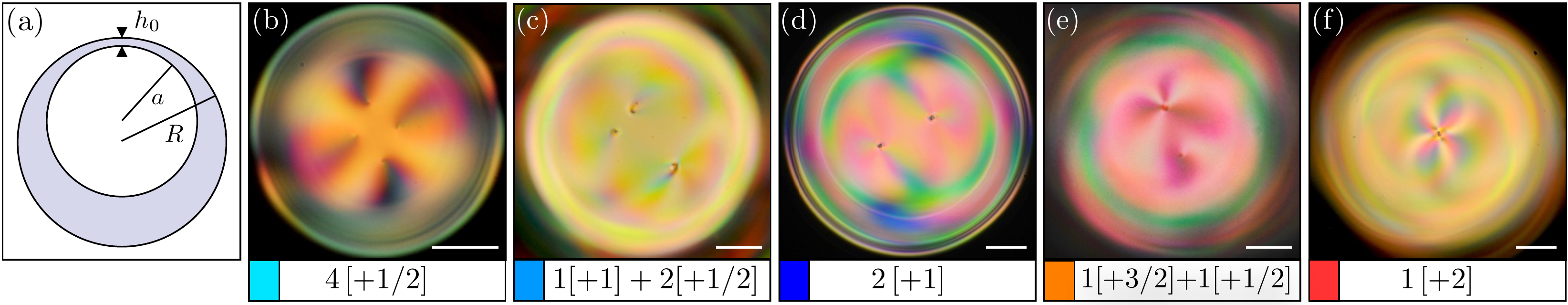}
 \caption{(a) Schematics showing a side view of a liquid crystal shell. (b-f) Top view of cholesteric liquid crystal shells between crossed polarisers. Each picture correspond to a specific defect configuration: (b) Four defects of winding number $+1/2$, (c) One defect of winding number $+1$ and two defects of winding number $+1/2$, (d) Two defects of winding number $+1$  \cite{Darmon2015}, (e) One defect of winding number $+3/2$ and one defect of winding number $+1/2$, (f) A single defect of winding number $+2$  \cite{Darmon2015}. Scale bar: $20\,\mu m$.}
 \label{Shells1}
\end{figure*}

In the present work, we study the new defect structures emerging in cholesteric shells for a wide range of shell thickness and cholesteric pitch. We show the existence of five possible configurations, provided that the molecules are tangentially anchored to the shell boundaries, which differ in the number and winding number of the defects. Interestingly, we report for the first time the existence of stable $+3/2$ defects in a spherical geometry. By looking at a very large sample of these shells, we show how these configurations are statistically distributed as a function of two relevant dimensionless parameters $u=(R-a)/R$ and $c=(R-a)/p$, where $a$ and $R$ respectively denote the inner and outer radii of the shell and $p$ is the cholesteric pitch. We study the detailed structure of each of the observed defects by bringing together experiments and numerical simulations, and show the existence of structures that are essentially different to those predicted for bulk droplets. We finally investigate the possibility of inducing transitions between defect configurations. By performing de-swelling experiments, we show that it is possible to induce topological transformations where the defects recombine themselves to form new defects of higher winding number. These transformations typically occur by following a well defined path. We finally study the intricate trajectories of the defects before recombining and develop a simple theoretical framework to explain the dynamics of these transitions.

\section{Equilibrium configurations}

\subsection{Experimental and numerical methods}

We use a glass capillary microfluidic device to generate cholesteric liquid crystal shells \cite{Utada2005}. The shells are double emulsions with the following composition: the inner and outer phases are composed of water with 1\%wt Polyvinyl Alcohol (PVA), and the middle phase is a mixture of 4-Cyano-4'-pentylbiphenyl (5CB) and a chiral dopant (S)-4-Cyano-4'-(2-methylbutyl)biphenyl (CB15). The amount of CB15 in the liquid crystalline solution determines the microscopic pitch, denoted $p$, of the resulting right-handed cholesteric helical arrangement  \cite{Ko2009}. The role of PVA is two-fold: $(i)$ it acts as a surfactant to stabilize the double emulsion and $(ii)$ it enforces planar degenerate anchoring on both inner and outer boundaries, meaning that the liquid crystal molecules are forced to lie tangentially to the two interfaces. The radii of the inner and outer droplets, see Fig.~\ref{Shells1}(a), are respectively denoted $a$ and $R$. In the present study, $R$ ranges between 30 and 90\,$\mu m$. The density mismatch between the inner aqueous solution and the liquid crystalline solution causes thickness heterogeneity in the shell. However, a disjoining pressure prevents contact between the two droplets, so that the minimal shell thickness is $h_0 \ne 0$  (see Fig.~\ref{Shells1}(a)). The average shell thickness can be defined as $h\equiv R-a$.  For each mixture, we ensure that we are far from the liquid crystal/isotropic phase transition to avoid defect nucleation or recombination, commonly observed close to the transition.

To gain insight into the detailed structure of the observed defects, we also perform numerical simulations. Since the shell thickness varies gradually and $R$ is large compared to $h$, the shell thickness gradient only affects the movement and the equilibrium position of the defects, but have negligible impact on their internal director structure. To show the structure of each defect, we thus assume a flat planar degenerate cell, which models a small area of the shell around the defect, and we enforce fixed winding number on the outer boundary. The simulation was done using the Landau-de Gennes free energy:
\begin{align}
  F=&\int_{\text{bulk}} \left\lbrace \frac{A}{2}Q_{ij}Q_{ji}+\frac{B}{3}Q_{ij}Q_{jk}Q_{ki}+\frac{C}{4}(Q_{ij}Q_{ji})^2 \right\rbrace {\rm d}V \nonumber \\
  +&\int_{\text{bulk}} \left\lbrace \frac{L}{2}Q_{ij,k}Q_{ji,k}+2q_0 L \epsilon_{ikl}Q_{ij}Q_{lj,k} \right\rbrace {\rm d}V\\
  +&\int_{\text{surface}} \left\lbrace \frac{W}{2}\left(\tilde{Q}_{ij}-\tilde{Q}_{ij}^\perp\right)^2\right\rbrace {\rm d}S \nonumber \ ,
\end{align}
which was then minimized with a finite difference method on a $360\times360\times200$ grid.  Note that the first two contributions respectively account for the phase transition and bulk elasticity, with $A$, $B$, $C$ the material parameters and $L$ the single elastic constant, consistently with previous studies  \cite{Sec2012,Darmon2015}. The auxiliary tensors $\tilde{Q}_{ij}$ and $\tilde{Q}_{ij}^\perp$ respectively denote the Q-tensor with added trace and its projection to the surface, as defined by Fournier and Galatola  \cite{Fournier2002}, and $q_0=2\pi/p$ is the intrinsic wave number of the cholesteric pitch. The last term in Eq.~(1) represents a surface anchoring term, where the anchoring strength was taken to be strong with $W=0.01\,{\rm J/m^2}$. The simulated slab thickness is $1.6{\,\rm \mu m}$. In each case, the initial condition was a pure $\chi$ cholesteric defect line with a chosen winding number, which was left to relax into the equilibrium structure. To highlight the symmetry and structure of both singular and nonsingular defects, we visualize them with the splay-bend parameter  \cite{Copar2013}.

\subsection{Defect configurations in cholesteric shells}

Because of the spherical nature of the interfaces delimiting the shell, any tangential nematic director field $\bm n$, where $\bm n$ represents the average molecular orientation, will be necessarily frustrated. Those frustrations are translated into topological defects which are singular points in the director field. Around those defects, the director experiences a $2 \pi m$ rotation, where $m$ is called the winding number. Since the symmetry of the nematic liquid crystal is only 2-fold, defect winding numbers can either be integers or semi-integers  \cite{Nelson2002}. The Poincar\'e-Hopf theorem \cite{Poincare1885, Hopf1926,Kamien2002} establishes that the winding numbers of the defects present on a surface must sum up to the surface Euler characteristic $\chi$, which in the particular case of a sphere equals $+2$. There are five different ways to satisfy this theorem using only positive winding numbers: i) One single $+2$ defect, ii) two $+1$ defects, iii) one $+3/2$ defect and one $+1/2$ defect, iv) one $+1$ defect and two $+1/2$ defects, and v) four $+1/2$ defects. Although all these configurations are compatible with the topological constrains, the configuration adopted by the shell will be, in principle, the one minimizing free energy.

\begin{figure}[t!]
 \centering
 \includegraphics[width=1\columnwidth]{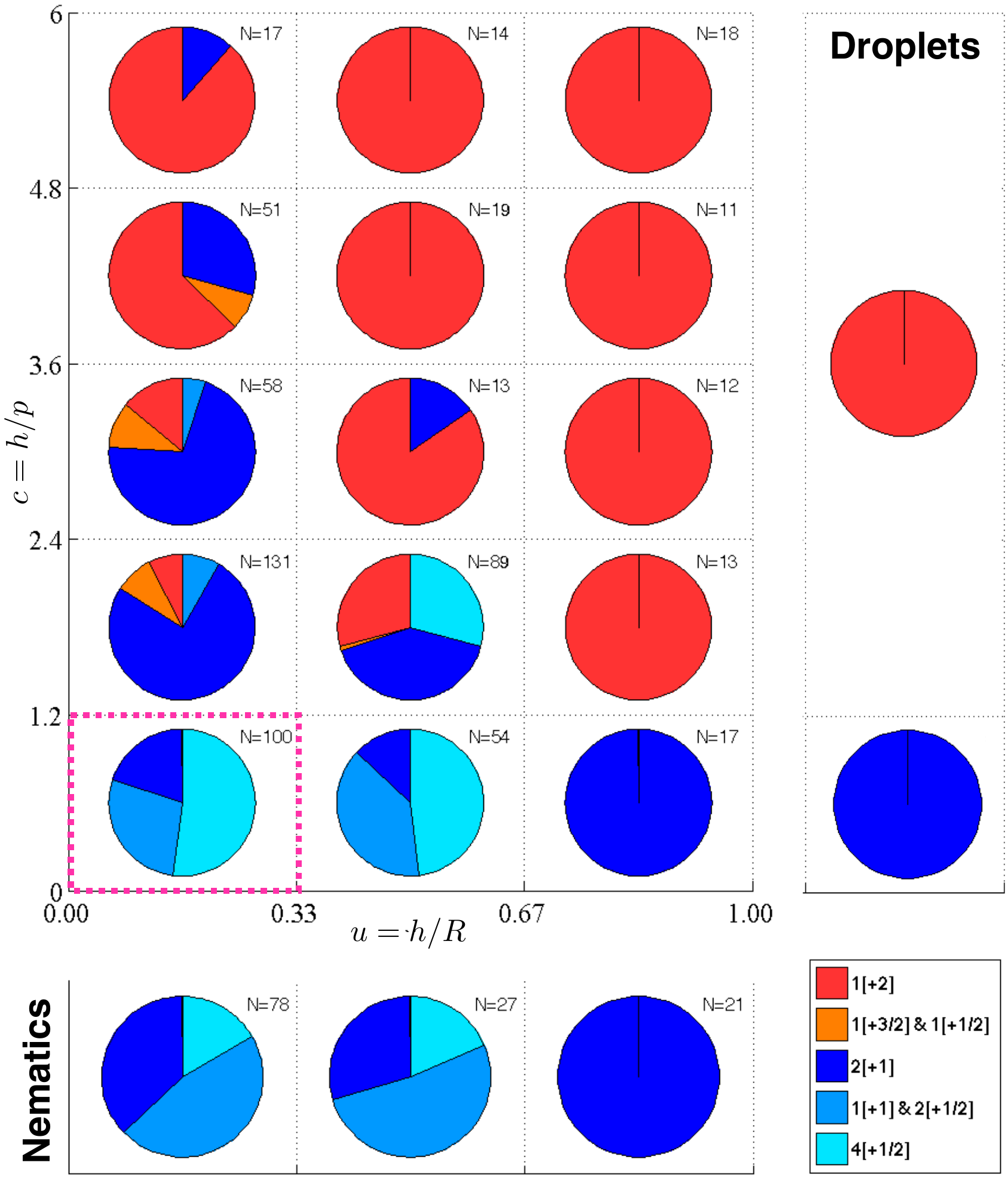}
 \caption{Statistical repartition of defect configurations in chiral nematic shells as a function of $u=h/R$ and $c=h/p$. The two limit cases, corresponding to nematic shells and cholesteric droplets, are respectively shown on the bottom and right sides of the diagram. The magenta dotted square is a visual help to refer to Fig.~\ref{tetra}. }
 \label{diagramme}
\end{figure}

\begin{figure}[b!]
 \centering
 \includegraphics[height=3.8cm]{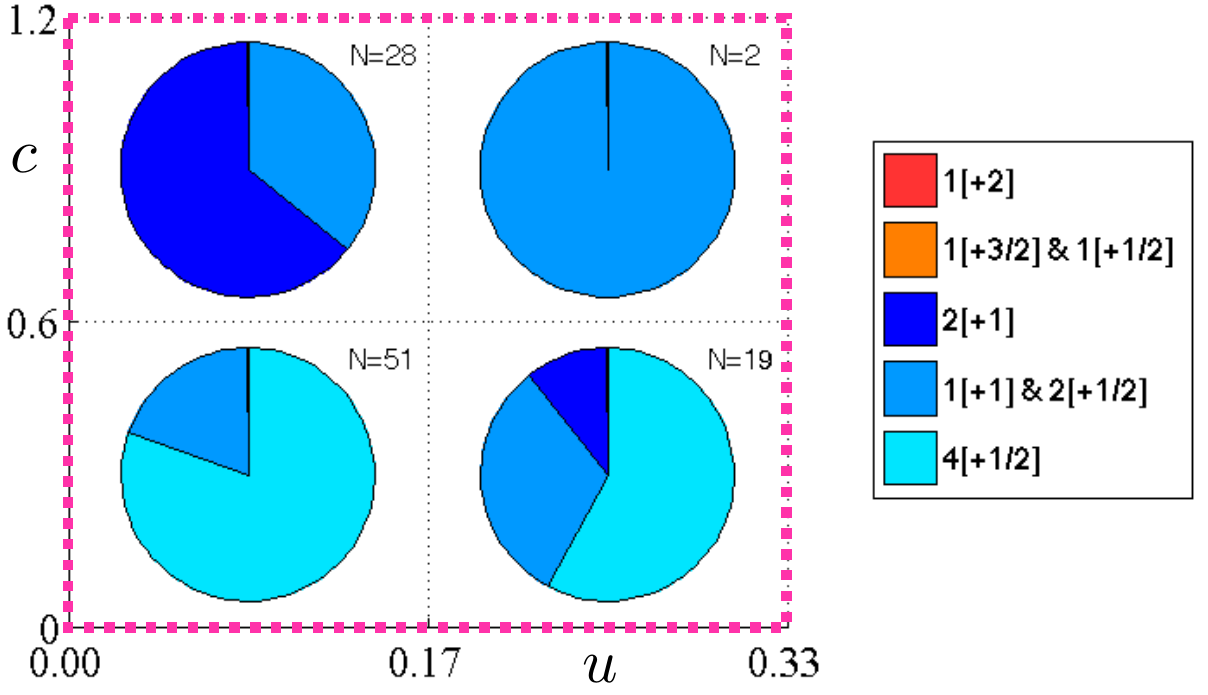}
 \caption{Statistical repartition of thin shells with little chirality as a function of $u=h/R$ and $c=h/p$. }
 \label{tetra}
\end{figure}

Three kinds of configurations have been reported for nematic shells  \cite{Fernandez2007,Lopez-Leon2011}. The first possible defect arrangement has four $+1/2$ defects. This defect configuration is the ground state for a purely two-dimensional nematic on a sphere  \cite{Lubensky1992,Nelson2002}. In the case of a shell, however, the defects are not surface point defects, but four singular disclination lines of winding number $+1/2$ that span the shell thickness. The second configuration is characterized by the presence of two $+1$ defects on each spherical surface. These surface defects, or boojums, associate into two pairs such that each defect on the outer sphere has its counterpart on the inner sphere. This defect configuration, which has an inherent three-dimensional character, is equivalent to the one observed in bulk nematic droplets. The subtle interplay between surface and bulk effects that takes place in shells becomes obvious in the third type of defect configuration observed, which is a hybrid state characterized by one $+1$ defect associated to two $+1/2$ defects  \cite{Lopez-Leon2011}. Hence, at the level of simple nematics, it is already clear that competition between surface and bulk effects plays a determinant role in the new type of defect configurations emerging in a shell geometry. This richness is expected to become even greater when inducing chirality in the nematic order.

When we add a chiral dopant to the nematic phase to produce a cholesteric shell, we indeed uncover a richer set of configurations, with a total of five different arrangements. These configurations are displayed in Figs.~\ref{Shells1} (b)-(f), which are cross-polarised images of the different types of cholesteric shell. In the images, the defects appear as dark points from which coloured brushes emerge. The number of coloured brushes, $M_i$, is related to the defect winding number, $m_i$, as $m_i=M_i/4$. The configurations shown in Figs.~\ref{Shells1} (b)-(d) are similar to those already observed in nematic shells, having four, three, and two defects, respectively. We also observe a configuration with a single $+2$ defect, see Fig.~\ref{Shells1} (f), which is characteristic of bulk cholesteric droplets  \cite{Bezic1992}. Finally, we observe a fifth and more intriguing configuration with one $+1/2$ defect and one $+3/2$ defect, see Fig.~\ref{Shells1} (e). This state was first theoretically imagined by Bezi\'{c} \& \v{Z}umer  \cite{Bezic1992} for cholesteric droplets but had never been observed before. The existence of stable $+3/2$ defects in a shell is itself remarkable, as they were only previously observed in specific planar cases  \cite{Madhusudana1982, Lee1982, MadhusudanaNV_MolCrystLiqCryst103_1983,LavrentovichOD_EurophysicsLetters12_1990,CuiL_LiqCryst26_1999, Li2003}. Interestingly, in cholesteric shells, all five possible configurations satisfying the Poincar\'e-Hopf theorem for positive winding numbers are found. In the following, and throughout the manuscript, we will use the notation $z_i\,[m_i] + z_j\,[m_j]$ to refer to the defect configurations, where $z_i$ denotes the number of defects with winding number $m_i$. 

In the shells shown in Fig.~\ref{Shells1}, the defects appear in the thinnest hemisphere of the shell, located either at the top or bottom of the shell depending on the sign of the density mismatch. Indeed, the equilibrium positions of the defects are ruled by a competition between an attractive force induced by the shell thickness gradient and a repulsive elastic defect interaction \cite{Lopez-Leon2011,Darmon2015}. It is worth mentioning that, in the $1\,[+1] + 2\,[+1/2]$ configuration, the outer defects sit at the vertices of a isosceles triangle with vertex angle $\alpha_0 \simeq 30^{\circ}$, see Fig.~\ref{Shells1}(b), regardless of the shell geometry. This cholesteric arrangement differs from its nematic counterpart, in which the triangle is not necessarily isosceles  \cite{Koning2015}. 

The elastic energies of the above configurations naturally differ from one another. To gain insight into the energy landscape associated to cholesteric shells, we look into the statistical repartition of each of these configurations. There are three characteristic length scales for cholesteric shells, namely the outer radius $R$, the inner radius $a$, and the cholesteric pitch $p$, from which two dimensionless parameters can be constructed. We select two meaningful parameters: $u=h/R$, which is a measure of the relative shell thickness, and $c=h/p$, called \textit{confinement ratio}, which counts the number of $2\pi$-turns of the molecular field over the average thickness of the shell, consistently with our previous study  \cite{Darmon2015}.

Fig.~\ref{diagramme} displays the statistical repartition of the five configurations for a number of shells $N_\text{tot}=743$, when varying $c$ between 0 and 6 and $u$ between 0 and 1. We measure $u$ and $c$ for each shell right after its creation, at rest, without any modification of its physico-chemical properties. The data are represented with pie charts, where the five different configurations are color-coded. The number of measured shells is indicated in each box. Note that the red and orange colors correspond to configurations found only in cholesteric shells. The two limit cases $c=0$ and $u=1$, corresponding respectively to nematic shells and cholesteric droplets, are also represented on the bottom and right parts of the diagram of Fig.~\ref{diagramme}. In the following we distinguish three cases, namely shells with large, intermediate and small thicknesses.

Thick shells,  \textit{i.e.} for $u \in [0.67,1]$, behave as droplets. At low chirality, \textit{i.e.} for $c<1.2$, only $2\,[+1]$ configurations are found, while at high chirality, \textit{i.e.} for $c>1.2$, the samples are only populated with $1\,[+2]$ configurations. This  tendency is exactly the same as the one observed in cholesteric droplets, for which there is a sharp transition between  $2\,[+1]$ and $1\,[+2]$ droplets at $R/p \simeq 1.2$  \cite{Lopez-Leon2011b}.

In shells with intermediate thickness, \textit{i.e.} for $u \in [0.33,0.67]$, confinement effects become more significant. For $c<1.2$, the three configurations reported in nematic shells are found  \cite{Lopez-Leon2011}. At $c=0$, the $1\,[+1] + 2\,[+1/2]$ configuration clearly dominates in the sample, although free energy calculations have shown that this arrangement is never the ground state of the system  \cite{Koning2015}.  When adding a little chirality, \textit{i.e.} for $c \in [0,1.2]$, we find the same three defect arrangements but with a notable difference in their statistical repartition. Indeed, a small but strictly positive confinement ratio seems to favor the $4\,[+1/2]$ configuration over the others. Interestingly, when increasing further the chirality in our samples,  \textit{i.e.} for $c \in [1.2,2.4]$, we observe that $(i)$ the $1\,[+1] + 2\,[+1/2]$ configuration disappears, $(ii)$ the relative populations of $4\,[+1/2]$, $2\,[+1]$ and $1\,[+2]$ are approximately equal and $(iii)$ there is a new configuration that seems to be specific to the cholesteric phase, namely the $1\,[+3/2] + 1\,[+1/2]$ configuration, although very rare (only one shell out of 189). For higher confinement ratios, \textit{i.e.} for $c>2.4$, the $1\,[+2]$ becomes largely predominant and eventually the only possible configuration for $c>3.6$.

Thin shells with low chirality, \textit{i.e.}  for $u \in [0,0.33]$ and $c<1.2$, are comparable to their intermediate counterparts in terms of statistical repartition of defect configurations, the only notable difference being that the proportion of $4\,[+1/2]$ is even larger in thin shells. As a matter of fact, a zoom on the lower left part of the diagram, displayed in Fig.~\ref{tetra}, reveals a remarkable feature. For very thin shells with little chirality, \textit{i.e.} for $c \in [0,0.6]$ and $u \in [0,0.17]$, the sample is populated mostly with $4\,[+1/2]$ shells (around 80\%). This could be particularly relevant in the context of colloidal self-assembly, since the $4\,[+1/2]$ configuration could be exploited to produce building blocks able to self-assemble into crystals with a diamond structure, which are expected to be perfect photonic band-gap materials  \cite{Nelson2002}.

The main differences between intermediate and thin shells occur at higher chirality. First, we see that the $2\,[+1]$ configuration represents a larger majority for $c \in [1.2,3.6]$. Second, we observe that the $4\,[+1/2]$ configuration disappears. Third, the hybrid $1\,[+3/2] + 1\,[+1/2]$ state becomes a non negligible part of the whole population. Finally, at very high confinement ratios, \textit{i.e.} for $c>3.6$, the $1\,[+2]$ configuration takes over the rest of populations. Hence, as it is often the case in physical systems, it is at the crossover regimes, in our case far enough from nematic shells and  cholesteric droplets, that the greater richness of configurations is found.


\smallskip
\section{Defect structures in chiral nematic shells}

\begin{figure}[t!]
 \centering
\includegraphics[height=8.5cm]{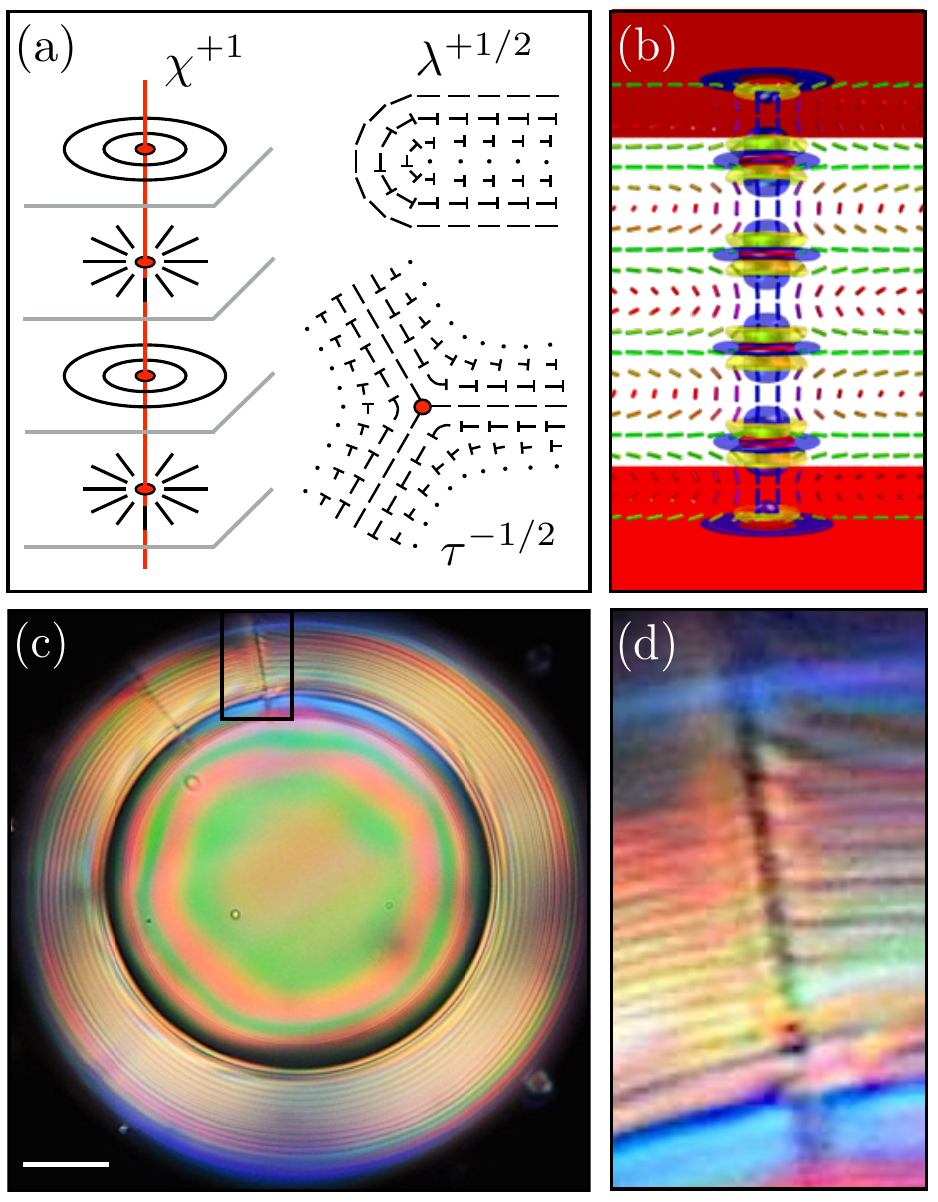}
 \caption{(a) Schematics of $\chi^{+1}$, $\tau^{-1/2}$ and $\lambda^{+1/2}$ disclinations in cholesterics. {(b) A simulated cross section of a $+1$ defect for $c=2.5$, showing that the defect core consists of a sequence of hyperbolic hedgehogs in the form of small $\tau^{-1/2}$ disclination rings, and a sequence of $\lambda^{+1/2}$ rings that terminate the layers. The splay-bend parameter  \cite{CoparS_LiqCryst40_2013} is used to highlight defects as regions of high deformation: blue and yellow regions respectively indicate zones of high splay and bend distortion.} (c) Side view of a $2[+1]$ shell between crossed polarisers, revealing a visible nonuniform structure of the defect core, which is enlarged in (d). Scale bar: $20\,\mu m$.}
 \label{DSSetRSS}
\end{figure}

Although nematic and cholesteric shells can be regrouped and compared in terms of defect winding numbers, cholesterics have an additional degree of order: the cholesteric twist axis. For this reason, the very nature of their disclinations is fundamentally more complex: in cholesteric liquid crystals, there are three possible types of disclinations called $\chi$, $\lambda$ and $\tau$, depending on whether the twist axis, the nematic director, or both, are singular. In a $\chi$ disclination line, the twist axis coincides with the line, where the director is singular, as shown in Fig.~\ref{DSSetRSS}(a). The $\tau$ and $\lambda$ disclination lines are characterised by a twist occurring perpendicularly to the disclination line, where the twist axis is singular. This is schematically shown in Fig.~\ref{DSSetRSS}(a), where  the nails represent an out-of-plane director field, with the nail heads indicating the direction at which $\bm n$ points upwardly. The $\tau$ disclination is also singular in terms of the director, whereas $\lambda$ has a non-singular core.

In a cholesteric shell with planar boundary conditions, the twist axis points perpendicularly to the surface everywhere except at the defects. Thus, all the defects have $\chi$ signature, with different semi-integer and integer winding numbers, when observing the surrounding director field far enough from their defect cores -- this feature is exploited in our simulations for an initial condition. However, it has been shown that cholesteric disclinations can relax locally in a non trivial fashion to minimize the free energy of the system  \cite{Bouligand1970,deGennes1993}. For example, we recently presented a detailed description of the intricate structure of the defects in the $2[+1]$ configuration of a cholesteric shell  \cite{Darmon2015}, see Fig.~\ref{DSSetRSS}(c) and Fig.~\ref{DSSetRSS}(d). We showed that, in addition to the two pairs of boojums appearing in the nematic case, here there is a number of alternating $\tau^{-1/2}$ and $\lambda^{+1/2}$ disclination rings that pile up though the shell connecting the upper and lower boojums of each pair. This structure is shown in Fig.~\ref{DSSetRSS}(b), where the dashed lines represent the director field. The defects can be identified by the blue and yellow isosurfaces, which indicate the regions of large splay and bend deformations, respectively. The singular rings, represented in red, are surrounded by regions of large splay elastic deformation.

\begin{figure*}[t!]
 \centering
 \includegraphics[width=1\textwidth]{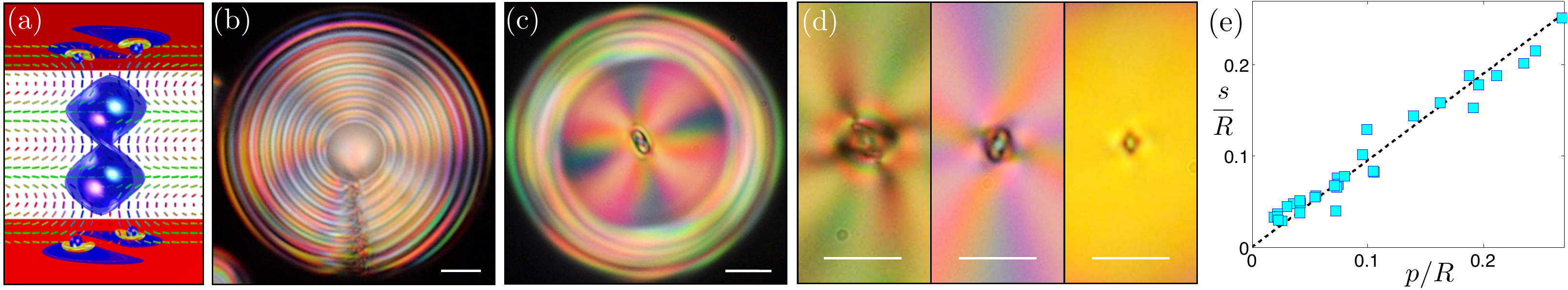}
 \caption{(a) Simulation of a nonsingular $+2$ defect core, consisting of two helically winding $\lambda^{+1}$ disclinations, ending as boojums at the boundary surfaces. The splay-bend parameter  \cite{CoparS_LiqCryst40_2013} is used to highlight defects as regions of high deformation: blue and yellow regions respectively indicate zones of high splay and bend distortion. (b) Side view of a $1[+2]$ shell between crossed polarisers. (c) Top view of a $1[+2]$ shell between crossed polarisers. (d) Crossed polarised images of +2 defects corresponding to shells with $p=9\,\mu$m, $p=3.6\,\mu$m, and $p=1.36\,\mu$m from left to right. (e) Rescaled defect spatial extension $s/R$ as a function of the rescaled cholesteric pitch $p/R$. Scale bar: $10\,\mu m$.}
 \label{RSScore}
\end{figure*}

Another non trivial disclination structure has been recently reported for the $1[+2]$ configuration \cite{Darmon2015}, see Fig.~\ref{RSScore}(b) and Fig.~\ref{RSScore}(c) for a top and side view of the shell. We showed that the disclination of global winding number $+2$ relaxes into two $\lambda^{+1}$ lines that wind around each other in a double-helix, as numerically predicted by \textit{Se\v{c} et al.}  \cite{Sec2012} for droplets, see Fig.~\ref{RSScore}(a). Two pairs of +1 boojums are also present on the inner and outer boundaries of the shell, which appear in Fig.~\ref{RSScore}(a) as two points of concentrated distortion at the upper and lower planes. An interesting feature concerns the size of the overall disclination structure, of total winding number $+2$, which seems to change with $p$. To investigate this, we consider $1[+2]$ shells obtained for different values of $p$. Fig.~\ref{RSScore}(d) shows three pictures of the defect cores, corresponding to $p=9\,\mu$m, $p=3.6\,\mu$m, and $p=1.36\,\mu$m from left to right. The scale bar is identical in each image and corresponds to $10\,\mu$m. All the pictures have been taken for very similar $R \simeq 50\,\mu$m. It is clear from Fig.~\ref{RSScore}(d) that the spatial extension $s$ of the defect structures increases with $p$.  More quantitatively, we even find that the spatial extent $s/R$ of the defect is directly proportional to the rescaled cholesteric pitch $p/R$, as shown in Fig.~\ref{RSScore}(e). 

\begin{figure*}[t!]
 \centering
 \includegraphics[height=4.5cm]{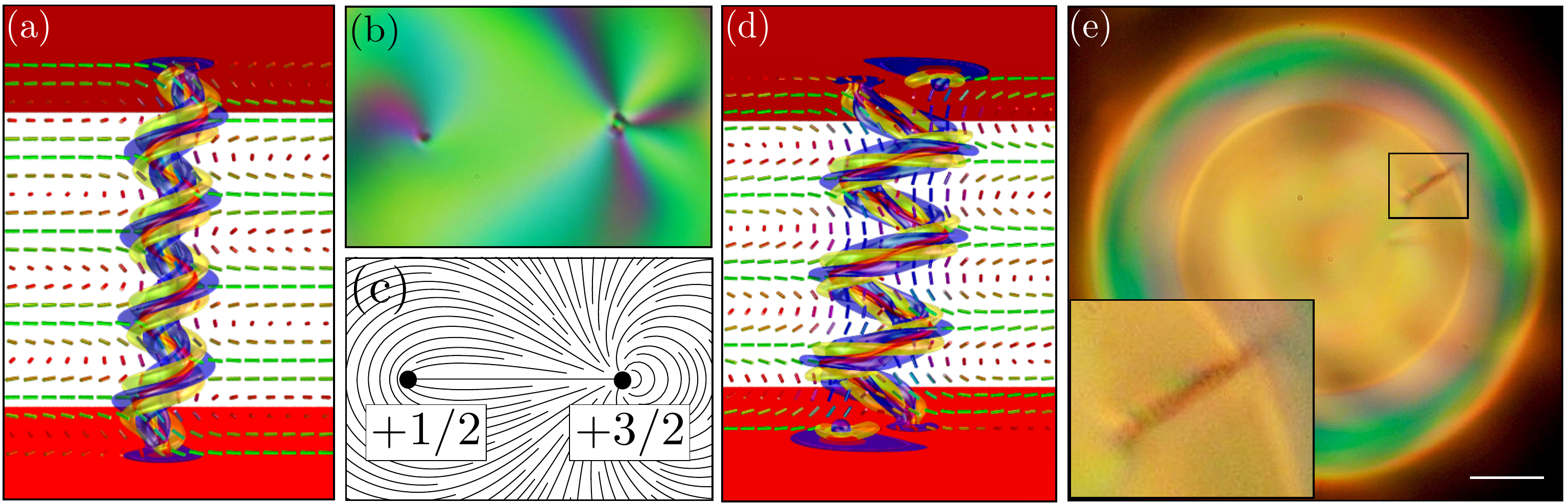}
 \caption{(a) A simulated $+1/2$ disclination line, which is locally composed of a helically shaped $\tau^{-1/2}$ singular core (with three-fold cross section, revealed by the splay-bend parameter) and a $\lambda^{+1/2}$ wound around it. (b) Cross-polarised image showing a top view of the $+3/2$ (right) and $+1/2$ (left) defects of a cholesteric shell. (c) Director field corresponding to the optical texture shown in panel (b). (d) A simulated $+3/2$ disclination core, composed from a more convoluted singular $\tau^{-1/2}$ line, wound around a nonsingular (escaped) $\lambda^{+1}$ line, which goes through the core and ends as two boojums at the surfaces. The entire structure is, as in panel (a), wrapped in a $\lambda^{+1/2}$ which terminates the layers. (e) Side view of a $1\,[+3/2] + 1\,[+1/2]$ shell between crossed polarisers. The inset shows a zoom in the $+3/2$ defect. Scale bar: $20\,\mu m$.}
 \label{Threehalf}
\end{figure*}

The first of the newly reported configurations in cholesteric shells is the tetravalent state characterised by four disclinations of $+1/2$ winding number. To investigate the nature of the observed $+1/2$ line, we perform numerical simulations. Instead of a pure straight $\chi^{+1/2}$ line, we see a singular disclination of helical shape with a period of half the cholesteric pitch, and a $\lambda^{+1/2}$ defect winding around it, terminating the cholesteric layers, see Fig.~\ref{Threehalf}(a). The singular disclination line has locally a $-1/2$ winding number, and resembles a $\tau^{-1/2}$ disclination, even though the twist axis is ill-defined around the core of the structure. The slope of the helix, together with the additional twist provided by the $\lambda$ disclinations, explain the seemingly contradictory transition from the $+1/2$ far-field winding and a $-1/2$ local winding of the singular defect core -- another demonstration that all singular disclinations in the director are topologically equivalent.

Another configuration that presents $+1/2$ defects is the  $1\,[+1] + 2\,[+1/2]$ configuration. The $+1$ defect resembles very much that of the $2[+1]$. Indeed, its larger spatial extent and very similar shape make us believe that it actually corresponds to the same structure. Similarly, the $+1/2$ defects seem to be identical in the $1\,[+1] + 2\,[+1/2]$ and $4\,[+1/2]$ configurations. The trivalent state can therefore be described as follows: one defect composed of alternating $\tau^{-1/2}$ and $\lambda^{+1/2}$ disclination rings, arranged as shown in Fig.~\ref{DSSetRSS} (b), and two $+1/2$ disclination lines with the structure shown in Fig.~\ref{Threehalf}(a).

The last but perhaps most intriguing defect combination is the state with $+3/2$ and $+1/2$ defects, see Fig.~\ref{Shells1}(d), which seems to be the first experimental evidence of stable $+3/2$ defects in cholesterics. The combination of a $+3/2$ and a $+1/2$ defect was imagined by \textit{Bezic et al.} \cite{Bezic1992} in their theoretical study of cholesteric droplets, but had never been observed before. Fig.~\ref{Threehalf}(b) and Fig.~\ref{Threehalf}(e) respectively show a top and side view of such defect configuration in an experimental cholesteric shell. According to the optical texture shown in Fig.~\ref{Threehalf}(b), the director field on the outer surface should be arranged as shown in Fig.~\ref{Threehalf}(c), where the $3\pi$ rotation of $\bm n$ around the $+3/2$ defect becomes evident. The side view of the $+3/2$ defect actually reveals the existence of a relatively thick line which appears to have a helical shape, see the inset in Fig.~\ref{Threehalf}(e).

We numerically investigate the inner structure of  the $+3/2$ defect by studying the relaxation of a $\chi^{+3/2}$ line. As in the case of a $+1/2$ defect, the core deforms into a helically twisted $-1/2$ singular disclination line, and a $\lambda^{+1/2}$ disclination terminating the regular layers, see Fig.~\ref{Threehalf}(d). However, due to additional winding that has to be compensated, there is another nonsingular $\lambda^{+1}$ going  through the center of the structure. This \textit{escaped} core has the director almost perpendicular to the shell surface, and ends as two boojums, just like in the $+1$ defect. Note that here, the $\lambda^{+1}$ does not decompose into a stack of small defect loops, but is wrapped tightly by the singular $-1/2$ disclination.

\section{Defects recombination and Lehmann effect}

We learned from the statistical study of cholesteric shells that the respective populations of defect configurations depend on both $u$ and $c$. In other words, changing the geometry and the confinement ratio of the shell influences the observed equilibrium configurations.  We recently showed that for cholesteric shells it is possible to transform a $2[+1]$ configuration into a $1[+2]$ configuration in a reversible way, by forcing the shell to move in the $u-c$ diagram  \cite{Darmon2015}. In this paper, we investigate the possibility of inducing transformations between other defect configurations. 

To change the shell parameters $u$ and $c$, we use osmosis.  By adding  CaCl$_2$ to the outer phase, we create a difference in osmotic pressure between the two aqueous phases that makes the inner droplet de-swell, resulting in the simultaneous increase of $u$ and $c$. We study the topological transformations undergone by shells having $4[+1]$, $1\,[+1] + 2\,[+1/2]$, and  $1\,[+3/2] + 1\,[+1/2]$ configurations. The de-swelling of a $4[+1]$ shell only make the defects become closer. As mentioned in Section 2, the equilibrium distance between defects results from a competition between an attractive force induced by the shell thickness gradient and a repulsive elastic defect interaction  \cite{Lopez-Leon2011,Darmon2015}. Therefore, when $u$ becomes larger, the shell becomes also more heterogeneous in thickness, shifting the equilibrium towards shorter defect distances. In a $1[+2]$  shell, when the two $+1$ defects are close enough, they come together and rearrange to form a single defect, so that the final state is the $1[+2]$ configuration. In a $4[+1]$ shell, however, the process ends differently. Indeed, we never observe a recombination of the defects, since the inner droplet is expelled from the shell when the defects become close enough. This actually means that the energy barriers associated to the possible transitions involving the $4\,[+1/2]$ configuration cannot be overcome by changing the geometry of the system. 

\begin{figure*}[t!]
 \centering
 \includegraphics[width=1\textwidth]{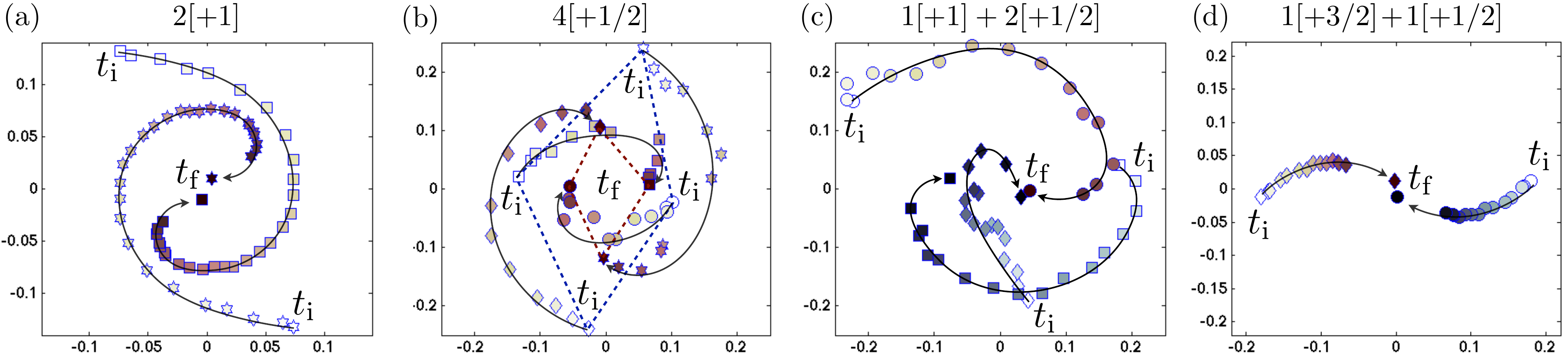}
 \caption{Defects trajectories in de-swelling experiments for the following defect configurations: (a) $2\,[+1]$, (b) $4\,[+1/2]$, (c) $1\,[+1] \, \& \, 2\,[+1/2]$, and (d) $1\,[+3/2] \, \& \, 1\,[+1/2]$.}
 \label{trajectories}
\end{figure*}

During the de-swelling process, we observe an interesting defect dynamics, where the defects get closer while turning around each other in what we called a defect \textit{waltz}, which we already reported for $1[+2]$ shells and explained as a result of a chemical Lehmann effect  \cite{Oswald2009}. Indeed, the radial current $\bm J$ of water molecules induces a torque $\bm \Gamma_\text{Leh}$ on the chiral molecules, provoking a rotation of the whole liquid crystal texture, and as a result, a rotation of the defects. This torque is related to the current through $\boldsymbol{\Gamma}_\text{Leh}= -\nu \bm J$, where $\nu$ is a phenomenological coefficient characteristic of the cholesteric mixture  \cite{Oswald2009}. The resulting defect trajectories for $4\,[+1/2]$ and $2\,[+1]$ shells are shown in Fig.~\ref{trajectories}(a) and (b), respectively. 

To investigate further possible transitions between configurations, we perform a de-swelling experiment in a $1\,[+1] + 2\,[+1/2]$ shell. As in previous experiments, the defects get closer as the shell de-swells. When they are close enough, the $+1$ defect fuse together with one of the  $+1/2$ defects, hence becoming a $+3/2$ defect, see the defect trajectories in Fig.~\ref{trajectories}(c). Nevertheless, we could not test further defect rearrangements in this experiment because the de-swelling process becomes very slow after a couple of hours. Indeed, the osmotic pressures in the inner and outer phases tend to equilibrate after some time, resulting in very slight changes of the shell geometry, hence losing the fuel for a possible transition. To check whether $+3/2$ and $+1/2$ defects are able to recombine, we perform a de-swelling experiment starting precisely from a shell with a $1\,[+3/2] + 1\,[+1/2]$ configuration. As shown in Fig.~\ref{trajectories}(d), $+3/2$ and $+1/2$ defects are indeed able to merge and form a single $+2$ defect. It is interesting to remark the $1\,[+1] + 2\,[+1/2]$ state can eventually evolve into a $1\,[+2]$ configuration, but by following a very specific path, where the $+1$ defect needs to recombine first with a $+1/2$ defect to form a $+3/2$ defect, which can in turn recombine with the remaining $+1/2$ defect to give rise to the final $+2$ defect. During all the de-swelling experiments, we observe a defect rotation similar to the one previously reported for $2\,[+1]$ shells. This can be explained by the fact that the Lehmann rotation depends neither on the nature nor on the number of defects present in the system. In all cases, we systematically find $\boldsymbol \Gamma_\text{Leh} \cdot \boldsymbol J > 0$, such that $\nu$ is always $<0$, as expected for a right-handed cholesteric, which is another good indicator that we are truly witnessing the Lehmann effect. 



\begin{figure}[b!]
 \centering
 \includegraphics[width=1\columnwidth]{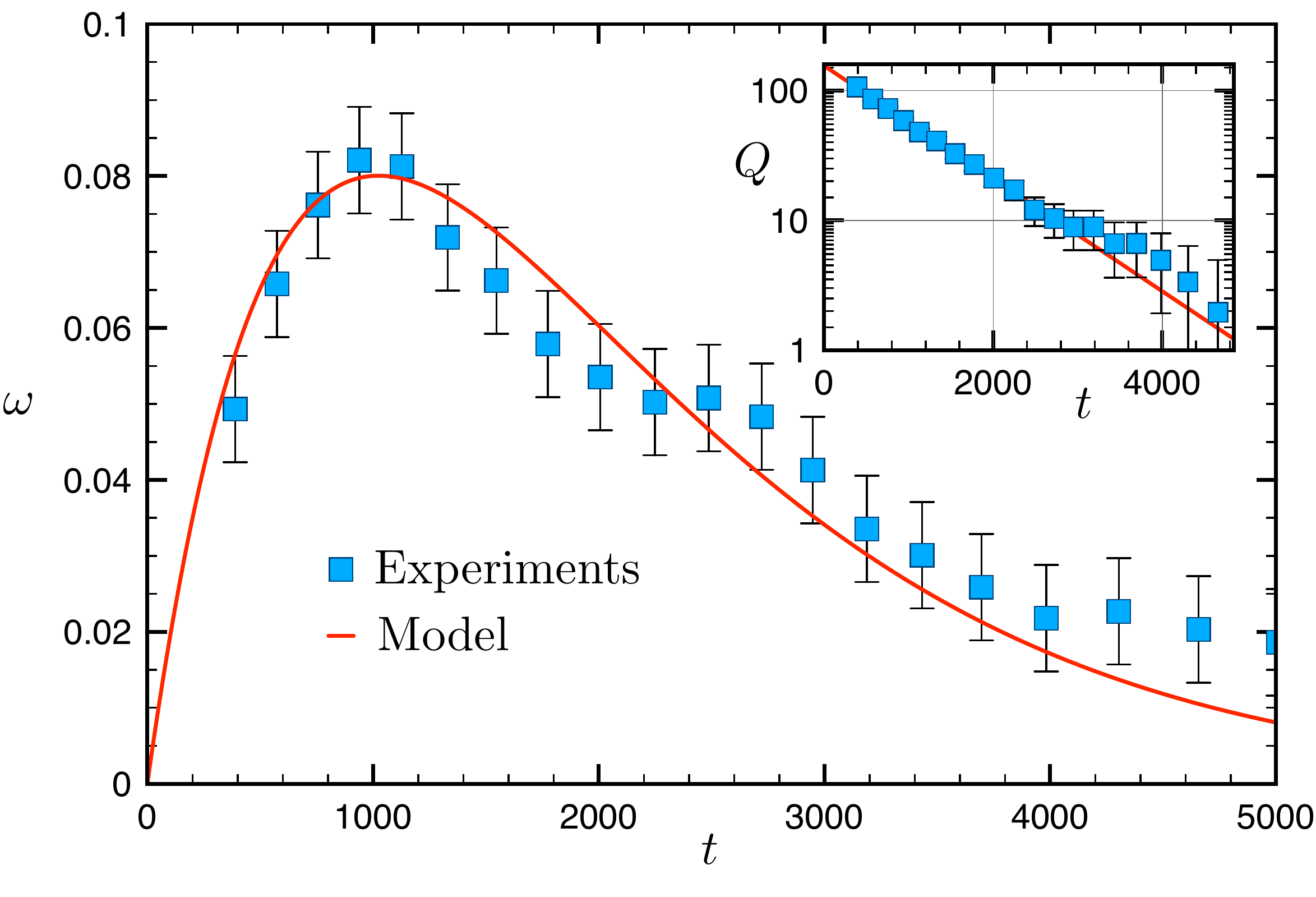}
 \caption{Angular rotation of the defects as a function of time in a typical de-swelling experiment, where the $2\,[+1]$ configuration evolves into the $1\,[+2]$ configuration.  Inset: Flow of water through the shell, $Q$, as a function of time. The blue squares correspond to the experimental data and the red line to the theoretical model.  The error bars correspond to the standard deviation of the rolling average performed on the experimental data.}
 \label{ModelLehmann}
\end{figure}

We wish to go a step further in the description of the Lehmann rotation by introducing a simple yet insightful theoretical framework. As mentioned previously, the chiral molecules of the cholesteric liquid crystal experience a torque $\boldsymbol \Gamma_\text{Leh}$, originating from the chemical potential gradient $\boldsymbol \nabla \mu$, itself related to $\boldsymbol J$ through $\boldsymbol J = -  \boldsymbol \nabla \mu$.  Considering then the liquid crystal as a permeable membrane of permeability $\xi$, one can relate the water flow $Q$ to the difference in chemical potential $\Delta \mu$ through $Q=\xi \mathcal A \Delta \mu / v_\text a$, where $\mathcal A$ is the area of the membrane and where $v_a$ is the molar volume. Noting finally that $\nabla \mu = \Delta \mu / h$, the Lehmann torque can be written as:
\begin{eqnarray}
\Gamma_\text{Leh}=\frac{\nu v_\text a}{\xi h \mathcal A} Q \ . 
\end{eqnarray}
Interstingly, while $h$ and $\mathcal A$ are both function of time, the product $h\mathcal A$ is not since it approximately corresponds to the volume of liquid crystal which is a conserved quantity throughout the experiment. As a result, only $Q$ is function of time in the Lehmann torque. Looking at the dynamics of such a system, one also needs to take into account  the viscous counter-torque $ \Gamma_\text{visc} = \eta \omega$, where $\eta$ is the bulk rotational viscosity and where $\omega$ is the angular velocity of the director field  \cite{Oswald2009}. In Fig.~\ref{ModelLehmann}, we plot the experimental angular velocity $\omega$ of the defects as a function of time (blue squares). As one can see, $\omega$ is time dependent and two parts can be identified in its evolution: $(i)$ it first increases and reaches a maximum, and $(ii)$ it  decreases on a time scale that is larger than that of the first ascending part. A first approach would naturally consist in balancing the two torques  \cite{Oswald2009}, yielding $\omega (t) \sim \frac{\nu v_\text a}{\eta \xi h \mathcal A} Q(t)$, and check whether $\omega$ and $Q$ indeed have the same temporal dependence. In the inset of Fig.~\ref{ModelLehmann}, we plot the water flow $Q$ as function of time, obtained from measuring how much the inner droplet de-swells during the experiment, on a log-lin scale. We see that Q monotonously decreases with time. 
The above-mentioned balance is therefore insufficient to describe the more complex behavior of $\omega(t)$. We thus need to add the observed transient regime to the theoretical framework, corresponding to the increasing part of $\omega(t)$. 
We do so through the following governing equation:
\begin{eqnarray}
\alpha \frac{\text d \omega}{\text d t}=\Gamma_\text{Leh} - \Gamma_\text{visc} \ , 
\label{equadiff1}
\end{eqnarray}
where $\alpha$ is an effective coefficient related to the transient regime. There is indeed a certain time for the osmotic pressure difference to be established, which is $\simeq 10^3\,$s for our system, according to Fig.~\ref{ModelLehmann}. Equation~\eqref{equadiff1} can be rewritten as:
\begin{eqnarray}
\tau_{\eta} \frac{\text d \omega}{\text d t} + \omega(t)= \beta Q(t) \ , 
\label{equadiff2}
\end{eqnarray}
where $\tau_{\eta}=\alpha/\eta$, and where $\beta=\nu v_\text a/(\eta \xi h \mathcal A)$. From the time evolution of $Q$ in the inset of Fig.~\ref{ModelLehmann}, it appears $Q$ is exponentially decreasing with time. In the following, we will therefore consider that $Q(t) = Q_0 e^{-t/\tau_Q}$ with $\tau_Q=1000\,$s, represented by the solid red line in the inset of Fig.~\ref{ModelLehmann}. The solution $\omega(t)$ to Eq.~\eqref{equadiff2} then reads:
\begin{eqnarray}
\omega(t)=\frac{\beta \tau_Q Q_0}{\tau_Q-\tau_{\eta}} \left(e^{-t/\tau_Q} - e^{-t/\tau_\eta}  \right) \ . 
\label{equadiff3}
\end{eqnarray}
This theoretical solution of $\omega(t)$ is displayed as a solid red line in Fig.~\ref{ModelLehmann}, with the best possible adjustable parameters. We find a rather good agreement between the data and our model, at least at a qualitative level, with $\tau_\eta=500\,$s. The small oscillations in the decreasing part of $\omega_\text{exp}(t)$  are probably an experimental artefact due to possible flows within the sample. 
These flows are constantly changing the local concentration of salt in the outer solution, which results in irregular osmosis dynamics. Note that there is also a small discrepancy between the model and the data at longer times,  due to the fact that the evolution of $Q$ is not strictly exponential (see inset Fig.~\ref{ModelLehmann}). Hence, our model seems to capture well the essence of the observed phenomenology, namely the faster inertial ascending part of $\omega(t)$, and the slower decrease following the decreasing water flow.


\section{Conclusions}
We provided a thorough study of the defect configurations appearing in cholesteric liquid crystal shells. We showed that five types of configurations are possible, revealing the greater richness of cholesteric shells as compared to their nematic counterparts. A remarkable result is the observation of stable $+3/2$ defects, which had only been observed before in exotic nematics or intricate confinements. Numerical simulations proved very efficient in gaining insight into the complex nature of the topological defects observed, which were composed by several disclination lines assembled into higher order structures. The formation of a given defect configuration depends on two dimensionless parameters, $c=h/p$ and $u=h/R$, where $h$, $R$ are the shell thickness and outer radius, respectively, and $p$ is the helical cholesteric pitch. By playing with these two parameters, we were able to induce transitions between configurations. In the allowed transitions, the defects approach each other by following intricate paths and an intriguing dynamics, which can be explained in terms of the chemical Lehmann effect.

\section*{Acknowledgements}

We thank S. \v{Z}umer, D. Se\v{c}, and A. Fernandez-Nieves for fruitful exchanges. We acknowledge support from the French National Research Agency by Grant 13-JS08-0006-01 and the Institut Pierre-Gilles
de Gennes (Laboratoire d’excellence, Investissements d’avenir, Program ANR-
10-IDEX 0001-02 PSL and ANR-10-EQPX-31).  S. \v{C}opar acknowledges support from Slovenian Research Agency under grants P1-0099 and Z1-6725.






\bibliographystyle{h-physrev}
\bibliography{biblio}

\end{document}